\documentclass{article}

\PassOptionsToPackage{numbers}{natbib}
\usepackage[preprint]{IEEE}
\usepackage{graphicx}
\usepackage{amsmath,amssymb,amsfonts}
\usepackage{subfig}
\usepackage{makecell}

\usepackage[utf8]{inputenc} 
\usepackage[T1]{fontenc}    
\usepackage{hyperref}       
\usepackage{url}            
\usepackage{booktabs}       
\usepackage{amsfonts}       
\usepackage{nicefrac}       
\usepackage{microtype}      

\usepackage{enumitem}

\usepackage[linesnumbered,ruled,vlined]{algorithm2e}
\usepackage{wrapfig}
\usepackage{lipsum}

\usepackage{multirow}
\usepackage{booktabs} 
\usepackage{array}    
\usepackage{geometry} 
\usepackage[table]{xcolor} 

\usepackage{enumitem}
\usepackage{graphicx}
\usepackage{subcaption}

\usepackage{array}

\usepackage{float}
\usepackage{placeins}
\usepackage{hyperref}

\usepackage{titletoc}
\usepackage{varwidth}
\usepackage{graphicx} 
\usepackage{lineno}

\title{ChatModel: Automating Reference Model Design and Verification with LLMs}

\author{%
\parbox{\textwidth}{\centering
Jianmin Ye\textsuperscript{1,2},
Tianyang Liu\textsuperscript{1,2},
Qi Tian\textsuperscript{1,2},
Shengchu Su\textsuperscript{1,2},
Zhe Jiang\textsuperscript{1,2},
Xi Wang*\textsuperscript{1,2}
\\ \textsuperscript{1}National Center of Technology Innovation for EDA, China \\
\textsuperscript{2}School of Integrated Circuits, Southeast University, China \\
}}

\begin{document}

\maketitle
\vspace{-2em}

\begin{abstract}
As the complexity of integrated circuit designs continues to escalate, the functional verification becomes increasingly challenging. 
Reference models, critical for accelerating the verification process, are themselves becoming more intricate and time-consuming to develop. Despite the promise shown by large language models (LLMs) in code programming, effectively generating complex reference models remains a significant hurdle. Therefore, we introduce ChatModel, an LLM-aided agile reference model generation and verification platform. ChatModel streamlines the transition from design specifications to fully functional reference models by integrating design standardization and hierarchical agile modeling. Employing a building-block generation strategy, it not only enhances the design capabilities of LLMs for reference models but also significantly boosts verification efficiency. We evaluated ChatModel on \textbf{300} designs of varying complexity, demonstrating substantial improvements in both efficiency and quality of reference model generation. ChatModel achieved a peak performance improvement of \textbf{58.99\%} compared to alternative methods, with notable enhancements in generation stability, and delivered a \textbf{9.18$\times$} increase in its capacity to produce reference model designs. 
Moreover, ChatModel accelerates the reference model design and validation cycles by an average of \textbf{7.11$\times$} over traditional manual approaches.
These results highlight the potential of ChatModel to significantly advance the automation of reference model generation and validation. 
\end{abstract}

\section{Introduction}
\label{sec:INTRO}

\begin{figure*}[t]
    \centering
    \begin{minipage}{0.442\textwidth}
        \centering
        \includegraphics[width=\textwidth]{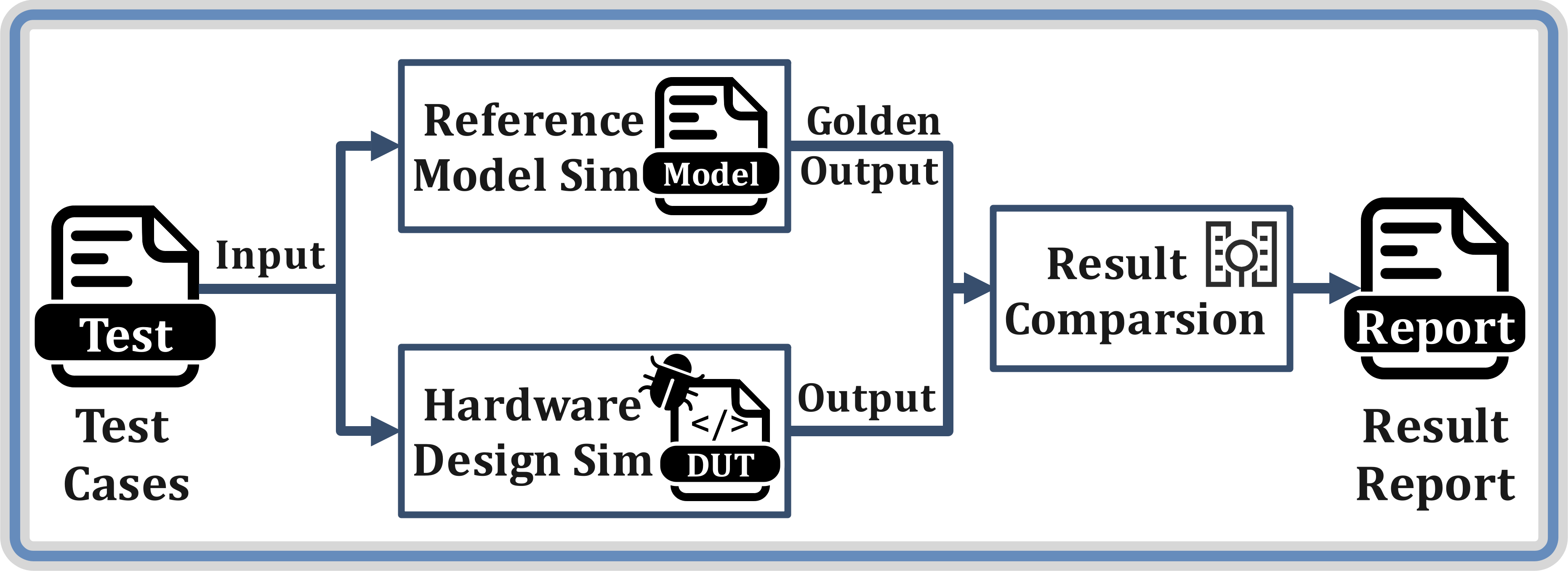}
           \caption{Model-Driven Verification Flow }
        \label{fig:MDV}
    \end{minipage}\hfill
    \begin{minipage}{0.552\textwidth}
        \centering
        \includegraphics[width=\textwidth]{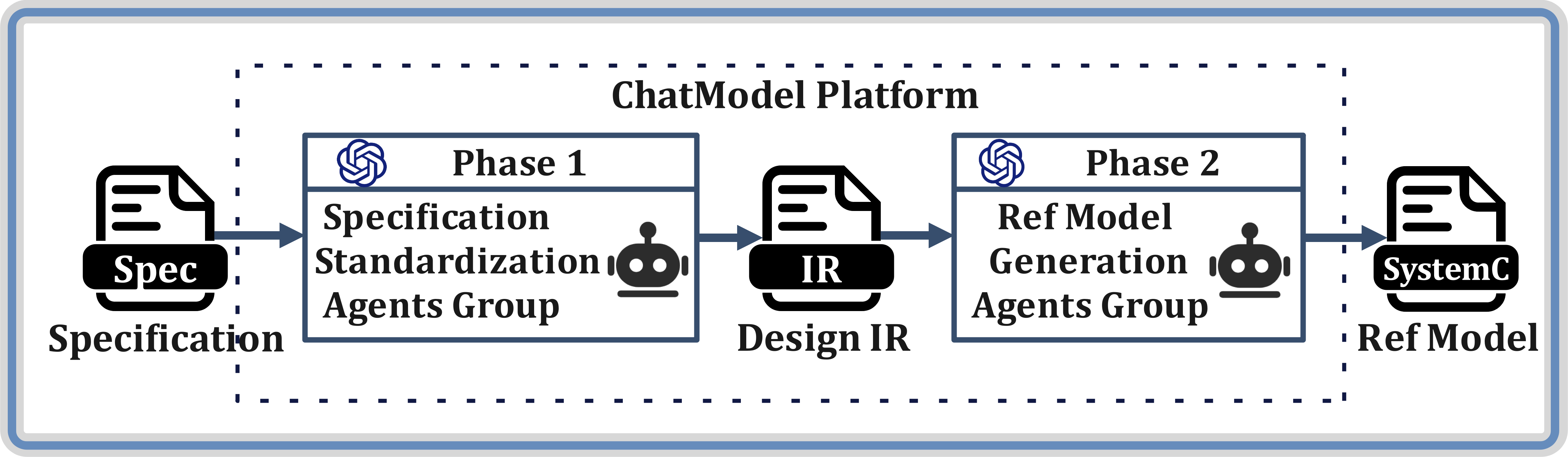}
        \caption{The Workflow of ChatModel}
        \label{fig:Framework}
    \end{minipage}
\end{figure*}

The rapidly increasing complexity of integrated circuit (IC) designs has significantly prolonged functional verification cycles, posing a critical bottleneck in agile hardware development~\cite{lahti2018we,bottleneck2,logicfuzzer}. 
In this context, model-driven verification (MDV) is recognized as a highly effective methodology and has been widely applied in preliminary chip designs, such as neural processors and artificial intelligence accelerators~\cite{quinones2020ampere,cirstea2024digital,mishty2024ai,linehan2012aspect,genius2017model,kaja2024advanced}. 
The models in MDV are typically written in high-level abstraction languages (i.e. Python~\cite{jiang2020pymtl3,batten2018open,logaras2014python}, C++~\cite{shcherbakov2011bringing,raia2024case}, MATLAB~\cite{delva2008using,versen2016model}, or SystemC~\cite{bergeron2012writing,atac2014hdl,mathur2007design,balasubramanian2023bit}) by engineers and serve as golden references for hardware designs. As shown in Figure~\ref{fig:MDV}, by systematically comparing the model outputs with those of the Device Under Test (DUT) under identical input stimuli, developers can effectively identify design flaws and accelerate the debugging and iteration process~\cite{mohamed2016ip}.

In the MDV framework, open-source simulators or loosely-timed modeling approaches are often employed to rapidly develop models~\cite{bellard2005qemu,ghosh2015case,bidmeshki2017information,becker2015challenges}. However, the execution processes within these models exhibit substantial discrepancies compared to actual hardware execution, thereby restricting their effectiveness in diagnosing errors within the DUT.
For the development of specialized designs, developers must consider timing factors to construct more accurate models.
However, building high-fidelity reference models requires a deep understanding of the behavior and verification requirements of the target hardware. 
As hardware designs grow increasingly intricate, the expertise and effort required expand exponentially, hindering agile model development.

With the advent of large language models (LLMs), a revolutionary paradigm shift has occurred in the field of natural language processing (NLP)~\cite{liu2023evaluating}. Recent studies have showcased their impressive capabilities in code generation for programming languages~\cite{xu2022systematic,ni2024l2ceval,jiang2024survey,liu2023your,wang2023review,zheng2024towards} such as C++ and Python, as well as hardware description languages~\cite{wang2024chatcpu,danneels2024exploring,wang2024large,wan2025genben} such as Verilog. However, their application in constructing reference models remains underexplored. General-purpose LLMs encounter numerous challenges when applied to reference modeling, including context and token limitations, hallucinations, and limited self-correction capabilities. These issues undermine the quality and stability of the outputs, restricting LLMs to small-scale designs. 
Moreover, the increasing diversity of design specifications and hardware architectures further exacerbates these issues, making it challenging for LLMs to meet the demands of constructing reference models.

To overcome these limitations, we introduce \textbf{ChatModel}, an end-to-end platform designed for agile generation and verification of reference models, as illustrated in Figure~\ref{fig:Framework}. 
The ChatModel leverages a multi-agent LLM system comprising two integrated groups: one standardizes design specifications, while the other automates the generation and validation of reference models. By adaptively decomposing complex design tasks into specialized subtasks, ChatModel facilitates the rapid construction of the corresponding design Intermediate Representations (IR). The Design IR serves as the input to the Hierarchical Agile Modeling (HAM) flow to generate the reference model. By employing an adaptive task planning algorithm, HAM facilitates automated and scalable generation and verification of reference models, seamlessly progressing from module to subsystem to system in a ``building block'' manner.

\begin{itemize}
    \item We propose the \textbf{ChatModel}, the first multi-agent platform for automated design and verification of reference models in SystemC, enabling agile hardware verification.
    \item We propose the \textbf{Design IR}, a novel domain-specific language specifically designed for LLM to accurately generate reference models, which effectively transforms complex specifications into a clear directed acyclic graph.
    \item We introduce a \textbf{Hierarchical Agile Modeling (HAM)} flow that realizes agile generation and verification of models through a modular generation methodology.
    \item We develop a benchmark, \textbf{ModelEval}, to comprehensively evaluate the capabilities of LLM in the generation of complex reference models.
    \item Our comprehensive experiments demonstrate that ChatModel outperforms other baseline methods, offering enhanced generative capabilities and more stable generations. Compared to general LLMs, it enhanced the LLM reference model generation ability by \textbf{10.96$\times$} and delivered an average \textbf{7.1$\times$} speedup in model generation for engineers compared to their manual approach.
\end{itemize}

The remainder of the paper is organized as follows:
Section~\ref{sec:Relatework} discusses related work. 
Section~\ref{sec:architecture} introduces the ChatModel platform.
Sections~\ref{sec:Standardization} \& ~\ref{sec:HAM} introduce the two phases of the ChatModel and some key components.
Section~\ref{sec:eval} evaluates ChatModel and analyzes the results.
Finally, Section~\ref{sec:con} concludes this paper.

\section{Related Works}
\label{sec:Relatework}

\textbf{Traditional Reference Model Design:} As a vital component in the verification process, reference models are typically developed using programming languages at varying abstraction levels, tailored to the functionality of the DUT~\cite{qamar2020comprehensive,moursi2018different}. For image or signal processing designs, Python~\cite{renuka2023monitoring,platonova2024using,beaulieu2017co} and MATLAB~\cite{lakshminarayana2022flattening,zheng2024generic} are commonly employed for modeling, while C++~\cite{heikura2021high,mintz2006hardware,raia2024case} is predominantly used for constructing reference models for logic functions and CPU designs. Recently, SystemC with Transaction-Level Modeling 2.0 (TLM 2.0) has achieved widespread adoption for reference model design, attributable to its capacity to facilitate modeling across multiple levels of abstraction~\cite{habibi2006design,jindal2003verification,park2008co,le2016towards,delbergue2016qbox}. For larger System-on-Chip (SoC) designs, simulators like gem5~\cite{binkert2011gem5,power2014gem5,ta2018simulating} or SST~\cite{rodrigues2011structural} are employed for model development to verify hardware system correctness, but their limited configurability results in coarse simulation granularity. Although these approaches have contributed to the advancement of reference model development, they require developers to possess substantial expertise in hardware and software programming, often necessitating a considerable and sustained investment of resources and time.

\textbf{LLM in Code Generation:} With the emergence of models such as ChatGPT~\cite{kocon2023chatgpt} and Claude~\cite{caruccio2024claude}, LLMs have demonstrated their powerful capabilities in automating code generation from natural language descriptions~\cite{coello2024effectiveness,zhang2023planning,zhong2024can,li2025idse}. Recent efforts~\cite{thakur2024verigen, nadimi2024multi, chang2023chipgpt, wang2024large, chang2024data,delorenzo2024make,wong2024vgv} have explored LLM-aided Hardware Description Language (HDL) generation and debugging using prompt engineering~\cite{d2024exploring,xu2025novel,niu2025rechisel,wan2024assessment}, retrieval-augmented generation~\cite{liu2024chatchisel,ping2025hdlcore}, and supervised fine-tuning~\cite{kumar2024hdl,yao2024hdldebugger,wu2024itertl}. Besides HDL generation, some works have also investigated complex software designs. The learning selection method LAIL~\cite{li2023large} and the self-planning code generation technique~\cite{jiang2023self} are both enhancements in these areas. A two-stage code generation method~\cite{zhao2024two} has demonstrated improved efficiency in generating code under zero-shot conditions. The LCG model~\cite{lin2024llm}, which adopts a multi-LLM agent cooperation mechanism, integrates seamlessly into traditional software development processes, significantly elevating the quality of generated code. In addition, study like UVLLM~\cite{hu2024uvllm} has explored the potential of LLM in the generation of reference models. However, implementations rely on the relatively lower-level language SystemVerilog, which restricts the scale of the generated models. Despite these advancements, the potential of LLMs to generate reference models for large-scale hardware designs remains underexplored, with notable limitations in both the token/context memory limits for complex designs and limited debugging capabilities of LLMs.

\section{Architecture \& Workflow}
\label{sec:architecture}

\begin{figure}[!t]
    \centering
    \includegraphics[width=1\linewidth]{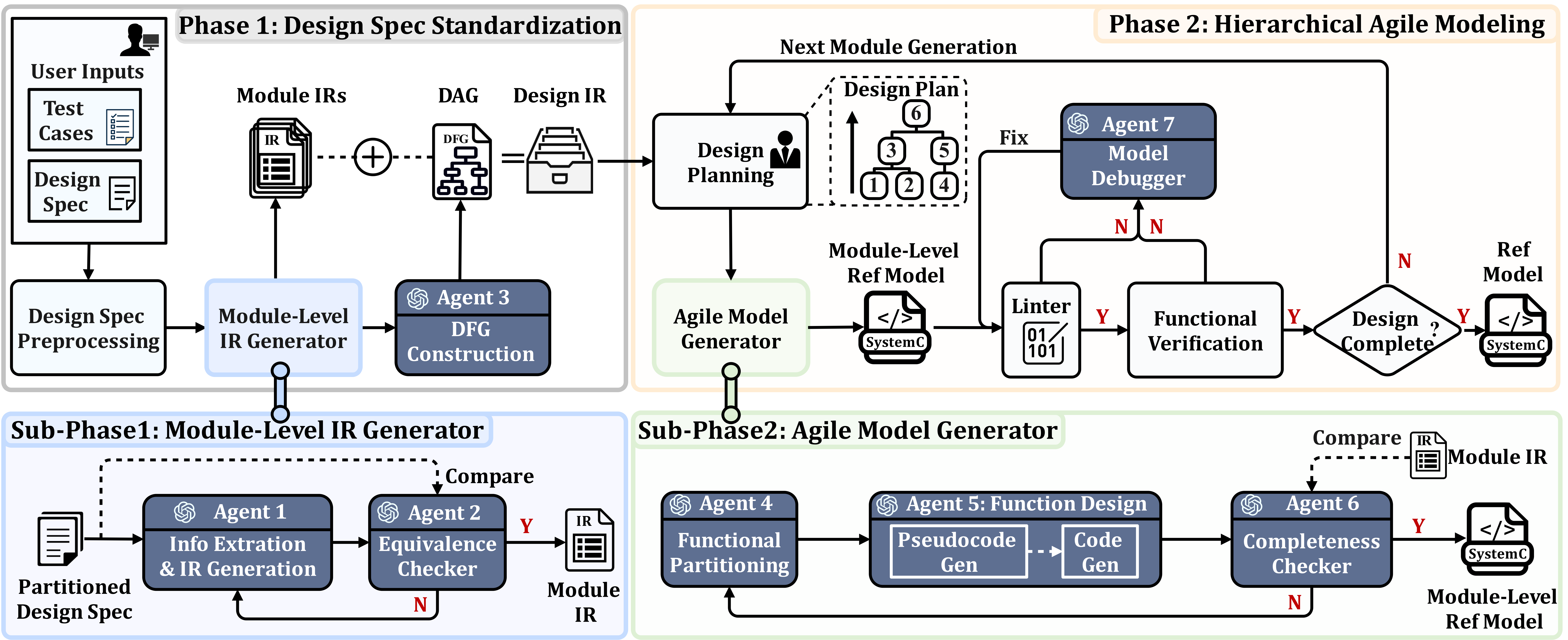}
    \caption{The workflow of the ChatModel Platform. In the first phase, the input design specifications are standardized into a design Intermediate Representation (IR). In the second phase, Hierarchical Agile Modeling (HAM) is employed to generate and verify reference models from module-subsystem-system based on the design IR.}
    \label{fig:chatmodel_workflow}
\end{figure}

Given that complex hardware designs consist of multiple functionally decoupled modules, ChatModel employs a ``divide-and-conquer'' approach to decompose intricate design tasks into manageable module-level subtasks, effectively addressing the aforementioned challenges presented in Section~\ref{sec:INTRO}. Building on this strategy, ChatModel offers a fully automated solution for spec-to-model generation using LLMs, as shown in Figure~\ref{fig:chatmodel_workflow}. This framework consists of the ``Design Specification Standardization Phase'' and the ``Hierarchical Agile Modeling Phase''. Each Agent is assigned a dedicated generation task and is equipped with a tailored chain of thought and template. 

\begin{figure}[t] 
    \centering
    \includegraphics[width=1\linewidth]{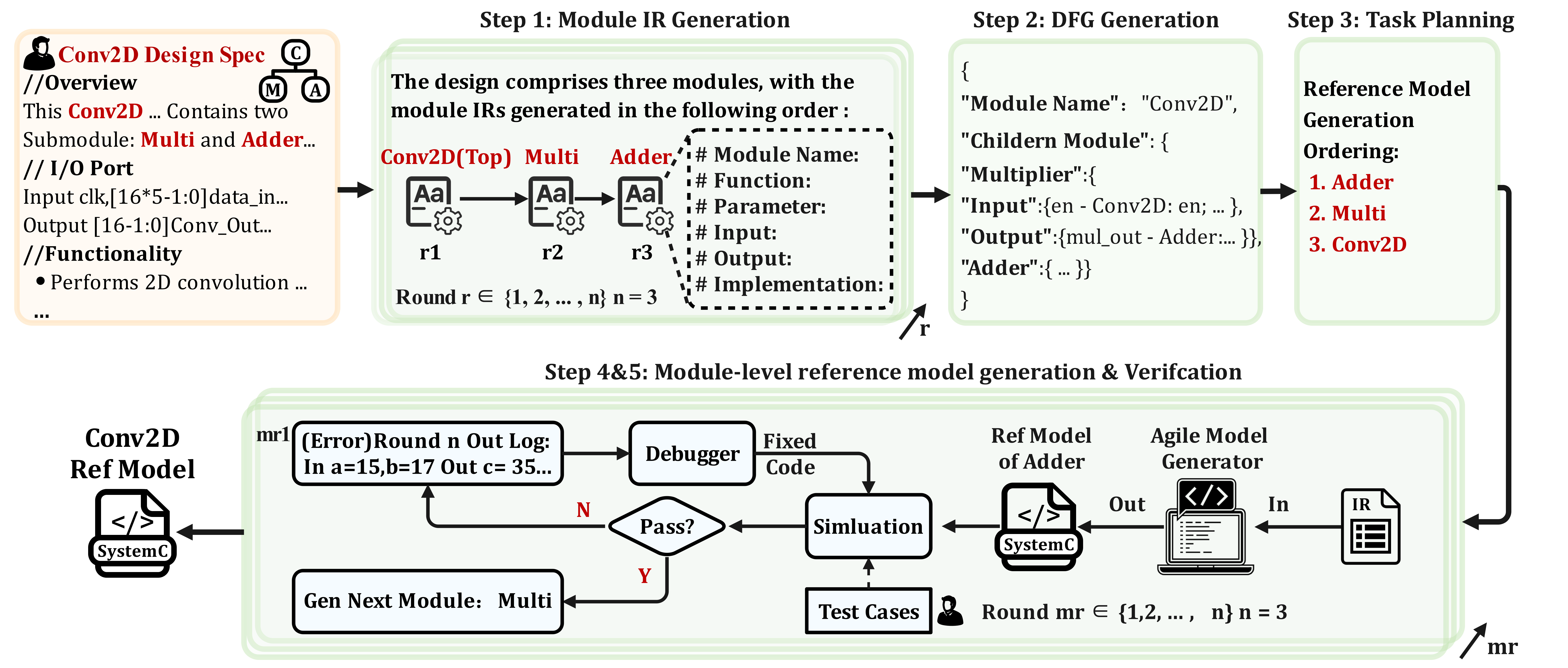}
    \caption{Examples of Reference Model Generation Flow. The design consists of a top-level module (Conv2D) and its two submodules (Multi and Adder). In Step 1, module-level IRs for Conv2D, Multi, and Adder are generated sequentially using a top-down approach. Conversely, the reference model is built by first generating and verifying the Adder and Multi submodules, then the Conv2D top module. The final reference model consists of these three files.}
    \label{fig:Example}
\end{figure}

The transition from design specifications to reference model generation involves five key steps. \textbf{(1)} \textit{Module IRs Generation (Agents 1 \& 2)}: Nonstandard input specifications are decomposed into multiple standardized module IRs based on the design structure, with each module IR corresponding to a specific generation subtask. \textbf{(2)} \textit{DFG Construction (Agent 3)}: For modules containing submodules, connections are established between parent module ports and internal submodule ports, effectively constructing the data flow graph. \textbf{(3)} \textit{Adaptive Task Planning}: Utilizing Algorithm~\ref{alg:hap}, the sequence for module generation and verification is defined, ensuring an organized and efficient workflow. \textbf{(4)} \textit{Module-Level Model Generation (Agents 4--6)}: Based on the module IR and DFG (optional), Algorithm~\ref{alg:amg} invokes the agile model generator to produce the corresponding reference model for each subtask. \textbf{(5)} \textit{Model Verification and Debugging (Agent 7)}: A comprehensive set of external test cases is employed to verify both the syntax and functionality of each generated module. Any module that fail verification are automatically repaired. If any modules remain incomplete after this process, subsequent modules are generated and validated iteratively until the top-level module is finalized, thereby completing the entire system design. To clearly demonstrate the workflow of ChatModel, Figure~\ref{fig:Example} illustrates the automatic generation and validation process of a reference model based on specific design specifications.


\section{Design Specification Standardization}
\label{sec:Standardization}

The content of a design specification can vary significantly with different authors and often becomes excessively verbose, which diminishes the ability of LLMs to accurately extract design-related information. To address these, we propose a design specification standardization flow that employs multiple agents to decompose design at the module level, extract implementation descriptions for each module, and construct inter-module connections. Through this process, the input design specifications are transformed into a Design Intermediate Representation (Design IR), which consists of multiple module IRs and a Design Architecture Graph (DAG).

The design specifications are first divided into multiple text segments through preprocessing, which serve as inputs for the subsequent module-level IR generator. \textbf{Module IRs} are generated in a top-down systematic manner, starting from the top module and sequentially producing each submodule. As depicted in Figure~\ref{fig:chatmodel_workflow}, when Agent 1 is tasked with IR generation, it first sequentially extracts port signals, implementation details, and other relevant information from the sliced specifications of the current module. This extracted information is then analyzed and refined, with irrelevant content filtered out and implementation methods optimized, resulting in the final module IR. Agent 2 performs equivalence checks between the generated module IR and the original design specification to assess whether any essential descriptive logic has been lost during the transformation process. For IRs that fail verification, detailed error reports are communicated to Agent 1, accompanied by targeted guidance to facilitate iterative refinement and accurate regeneration. By employing multiple rounds of reading slice content and self-checking, ChatModel effectively mitigates attention loss in LLMs when processing long texts.

After the successful verification of all module IRs, Agent 3 constructs the \textbf{DAG} by systematically analyzing the port data flow graph of parent and child modules. The DAG abstracts the hierarchical structure of hardware modules into a structured directed acyclic graph, effectively resolving interface alignment challenges inherent in modular design. It comprises multiple depth-2 subtrees, where each subtree delineates the port connections between a parent module and its child modules, providing a precise and structured reference for subsequent generation processes.

\section{Hierarchical Agile Modeling}
\label{sec:HAM}

\vspace{-0.5em}
In this section, we introduce the four key components of Hierarchical Agile Modeling: \textit{Design Planning}, \textit{Agile Model Generator}, \textit{Model Verification}, and \textit{Model Debugger}.
\vspace{-0.5em}

\subsection{Design Planning}

\begin{wrapfigure}{r}{0.60\textwidth}
    \vspace{-15pt}
    \hspace{4pt} 
    \begin{minipage}{0.59\textwidth}
    \begin{algorithm}[H]
    \caption{\textbf{Hierarchical Adaptive Planning}}
    \label{alg:hap}
    \KwIn{Design Architecture graph \( D = \{ d_1,\dots,d_n \} \)}
    \KwOut{Optimized Task Sequence \( T_{\text{Seq}} \)}
    \SetAlgoLined
    \SetKwFunction{FMain}{Build\_Seq}
    \SetKwFunction{FHAP}{HAP}
    \SetKwProg{Fn}{Function}{:}{}

    \Fn{\FMain{$D_{\text{tree}}$, node, visited}}{
        \If{node $\in$ visited}{
            \Return{$\varnothing$}
        }
        visited.add(node)\;
        \( C = \{ c \mid (p, c) \in D_{\text{tree}}, p = \text{node} \} \)\;
        \( C_{\text{norm}} = \text{NormalizeChildren}(C) \)\;
        \( seq = \varnothing \)\;
        \textit{// Post-order Traversal}\;
        \ForEach{$\text{c\_node} \in C_{\text{norm}}$}{
            \( seq = seq + \FMain(D_{\text{tree}}, \text{c\_node}, visited) \)\;
        }
        \( seq = seq + [node] \)\;
        \Return{seq}\;
    }

    \Fn{\FHAP{$D$}}{
        \textit{// Parent-Child Relation Extraction}\;
        \( D_{\text{tree}} = \varnothing \)\;
        \ForEach{$d_i \in D$}{
            \( (p_i, c_i) = \text{ExtractParentChildPairs}(d_i) \)\;
            \( D_{\text{tree}} = D_{\text{tree}} \cup \{(p_i, c_i)\} \)\;
        }
        \textit{// Root Identification}\;
        \( T_{\text{root}} = \text{IdentifyRoot}(D_{\text{tree}}) \)\;
        \textit{// Sequence Construction}\;
        \( \text{visited} = \varnothing \)\;
        \( T_{\text{Seq}} = \FMain(D_{\text{tree}}, T_{\text{root}}, \text{visited}) \)\;
        \Return{$T_{\text{Seq}}$}\;
    }

    \end{algorithm}
    \end{minipage}
    \vspace{-20pt}
\end{wrapfigure}

To meet the needs of generating reference models of various scales and agile verification of models, this work proposes a \textbf{Hierarchical Adaptive Planning (HAP)} algorithm to replace the traditional LLM random design task planning strategy. With HAP, ChatModel can automatically construct and execute the optimal task sequence even when the hardware design structure and scale change.

As shown in Algorithm~\ref{alg:hap}, this algorithm preprocesses the input DAG by sequentially extracting parent-child module pairs \((P_i, C_i)\) and storing them in the design tree \(D_{\text{tree}}\). Next, HAP identifies the root node \(T_{\text{root}}\) of \(D_{\text{tree}}\) and uses a post-order traversal strategy to traverse it. The algorithm explores deeper layers from the starting node, backtracking when no further exploration is possible, and continues along other unvisited paths. Since hardware designs often instantiate multiple modules with different names but the same functionality, several similar subtrees that do not require redundant generation may appear in \(D_{\text{tree}}\). Therefore, HAP optimizes the generation of task planning paths by normalizing node names to identify and bypass certain subtrees, thereby improving the efficiency of reference model generation. The search strategy of HAP ensures that dependent modules are ready before subsequent modules are generated, allowing the validation of the internal subsystems during the generation process rather than waiting for the entire design to be completed.

\subsection{Agile Model Generator}

The \textbf{Agile Model Generator (AMG)} is responsible for delivering a stable and high-quality reference model based on the input Module IR and DAG. Within AMG, three cooperative agents manage the refinement of generation tasks, SystemC code generation, and code completeness checks. This collaborative process ensures that the resulting reference models meet the specified requirements.

\begin{wrapfigure}{r}{0.60\textwidth}
    \vspace{0pt}
    \hspace{4pt}
    \begin{minipage}{0.58\textwidth} 
    \raggedright
    \begin{algorithm}[H]
    \caption{\textbf{Automated Model Generation}}
    \label{alg:amg}
    \KwIn{Module IR \( M_{\text{IR}} \), Module DAG \( \text{d}_i\) }
    \KwOut{Reference model \( Ref_{\text{Model}} \)}
    \SetAlgoLined
    \SetKwFunction{FAMG}{AMG}
    \SetKwProg{Fn}{Function}{:}{}
    \Fn{\FAMG{$M_{\text{IR}}$, $\text{d}_i$}}{
        // Functional block partition -- Agent 4\;
        \( f_{\text{blocks}} \gets \text{aggregate\_func\_blocks}(M_{\text{IR}}) \)\;
        \( \text{r}_{\text{c}} \gets \varnothing \)\;
        \If{module is parent module}{
            \( r_{\text{c}} \gets \text{analyze\_interactions}(M_{\text{IR}}, \text{d}_i) \)\;
        }
        // Model generation \& verification -- Agents 5\& 6\;
        \( \text{c}_{\text{result}} \gets \varnothing \)\;
        \While{$\text{c}_\text{result}$\text{ is not pass} }{
            \( \text{code}_{\text{h}} \gets \text{gen\_header}(M_{\text{IR}}, f_{\text{blocks}}, \text{r}_c, \text{c}_{\text{result}}) \)\;
            \ForEach{\( f_{\text{block}_i} \in f_{\text{blocks}} \)}{
                \( info_{\text{i}} \gets \{ M_{\text{IR}}, f_{\text{block}_i}, \text{code}_{\text{h}}, \text{r}_c, \text{c}_{\text{result}} \} \)\;
                \( \text{code}_{\text{s}_i} \gets \text{func\_impl}(info_{\text{i}}) \)\;
                \( \text{code}_{\text{s}} \gets \text{code}_{\text{s}} \cup \text{code}_{\text{s}_i} \)\;
            }
            \( \text{c}_{\text{result}} \gets \text{eq\_check}(\text{code}_{\text{h}}, \text{code}_{\text{s}}, M_{\text{IR}}, \text{d}_i) \)\;
        }
        \( \text{Ref}_{\text{Model}} \gets \{\text{code}_{\text{h}},\text{code}_{\text{s}}\} \)\;
        \Return{$Ref_{\text{Model}}$}\;
    }
    \end{algorithm}
    \end{minipage}
    \vspace{-15pt}
\end{wrapfigure}

As shown in Algorithm~\ref{alg:amg}, Agent 4 invokes the corresponding module IR \(M_{\text{IR}}\) based on the current generation task and employs CoT reasoning to analyze each functional point in the \(M_{\text{IR}}\). Similar functional points are grouped into the same functional block, forming a set of blocks denoted as \( f_{\text{blocks}} \), where each block corresponds to a single round of generation dialogue. This multi-round strategy ensures that the LLM focuses on generating one functional block per round, thereby minimizing incomplete outputs and hallucination errors caused by token limitations in LLMs. 
For parent modules, as certain child-module port signals may directly interact with the parent’s internal logic, Agent 4 refines the DFG to explicitly capture these interactions and stores the results in \( r_{\text{c}} \).

Subsequently, Agent 5 constructs the relevant module class declarations and process \( code_h \) based on the provided information. During the function design phase, Agent 5 employs a multi-turn dialogue approach to sequentially generate function code\( code_{\text{s}_i} \) for each functional block \( f_{\text{block}_i} \), thereby completing the implementation of the module. Introducing an intermediate step of pseudocode generation before formal function implementation effectively enhances the accuracy of LLM in generating reference models. Finally, ChatModel uses Agent 6 to verify the functionality, as well as the integrity and equivalence of port signals between the generated reference model and the original IR, ensuring the correctness of the model.

\subsection{Model Verification \& Debugger}

The validation strategy of ChatModel for the reference model adheres to the task planning generated by HAP. Upon generation of the reference model for a specific module, user-provided test cases are utilized to verify the syntax and functional accuracy of the reference model. The model is deemed functionally equivalent to the hardware design only after successfully passing all tests. Subsequent subsystem or system-level models will incorporate these validated models and undergo comprehensive subsystem or system-level validation. This incremental modular validation approach employs a bottom-up strategy, facilitating concurrent validation during the development of internal subsystems. In addition, it effectively reduces validation complexity, simplifies error localization, and accelerates convergence toward an accurate model.

To enhance the debugging capabilities of LLMs, we document common errors and their solutions within the reference model and utilize Chroma~\cite{prabhune2024deploying} to construct a vector database. Agent 7 employs Retrieval-Augmented Generation (RAG) technology to access relevant data from this database, facilitating the resolution of syntactic and functional errors during validation. For syntactic errors, terminal output logs are used to retrieve pertinent information from documents or knowledge bases via RAG technology. This information, along with relevant code snippets, is input into the LLM for error analysis and code correction. In the case of functional errors, Agent 7 also extracts logs from the period surrounding the error occurrence to enable the LLM to analyze the causes of output discrepancies.

\section{Evaluations}
\label{sec:eval}

In this section, we evaluate the capability of ChatModel in generating reference models for various designs across multiple dimensions. The results demonstrate that the framework exhibits exceptional performance and remarkable stability when processing diverse designs, effectively enhancing the capability and efficiency of LLMs in agile model generation and verification tasks.

\subsection{Experimental setup}
\noindent\textbf{Benchmarks.} We selected \textbf{300} hardware designs from the widely-used benchmarks VerilogEval2.0~\cite{pinckney2024revisitingverilogevalnewerllms} and RTLLM2.0~\cite{lu2024rtllm}, as well as our own ModelEval dataset. Although VerilogEval2.0 and RTLLM2.0 are primarily used to evaluate the HDL generative capabilities of LLMs, their design specifications can still be applied to generating reference models. 
To ensure that our evaluation closely reflects real-world design tasks, we introduce the ModelEval benchmark, which comprises 94 diverse designs representative of practical application scenarios. Each benchmark design is accompanied by comprehensive test cases to rigorously verify the functional correctness of the generated reference models.

\noindent\textbf{Evaluation Metrics.} Since LLMs are probabilistic, each case is evaluated multiple times to account for output variability, and the pass@k~\cite{chen2021evaluating} is calculated. Unless otherwise specified, the experiment uses GPT-4o as the foundational LLM for each agent via the OpenAI API. The temperature of all LLMs is set to 0.3 to balance the generation capability and randomness. The \textit{debugging iteration threshold} in ChatModel is set at \textbf{6}, as improvements beyond this point are difficult to observe.

\noindent\textbf{Baseline Methods.} Existing reference model construction methods still rely on manual development, and there is a lack of automated generation approaches. Therefore, we select several representative LLM-based baselines for comparison, including \textbf{Zero-Shot}~\cite{xian2018zero} (generation without examples), \textbf{CoT+Few-Shot}~\cite{kim2023cot} (step-by-step reasoning with examples), and \textbf{Multi-Agent}~\cite{zhao2024two} (one agent analyzes, and another generates).

\subsection{Capability for Reference Model Generation}
\label{sec:ref_model_generation}

\begin{table*}[t]
\caption{Pass@5 results of SystemC reference model generation by ChatModel and LLM-based methods across benchmarks and LLMs. \textbf{SPR} represents the Syntax Pass Rate, and \textbf{FPR} denotes the Functionality Pass Rate. For each LLM group and column, the highest score is in bold and the second highest is underlined.}
\centering
\small
\renewcommand{\arraystretch}{1.08}
\setlength{\tabcolsep}{4pt}

\begin{tabular}{|>{\centering\arraybackslash}m{2.55cm}|
                >{\centering\arraybackslash}m{1.05cm}|
                >{\centering\arraybackslash}m{1.05cm}|
                >{\centering\arraybackslash}m{1.05cm}|
                >{\centering\arraybackslash}m{1.05cm}|
                >{\centering\arraybackslash}m{1.05cm}|
                >{\centering\arraybackslash}m{1.05cm}|
                >{\centering\arraybackslash}m{1.05cm}|
                >{\centering\arraybackslash}m{1.05cm}|}
\hline
\rule{0pt}{10pt}
\multirow{2}{*}{\textbf{Method}} 
& \multicolumn{2}{c|}{\textbf{VerilogEval2.0}} 
& \multicolumn{2}{c|}{\textbf{RTLLM2.0}} 
& \multicolumn{2}{c|}{\textbf{ModelEval}} 
& \multicolumn{2}{c|}{\textbf{Average}} \\
\cline{2-9}
\rule{0pt}{14pt}
& SPR/\% & FPR/\% & SPR/\% & FPR/\% & SPR/\% & FPR/\% & SPR/\% & FPR/\% \\
\hline

\multicolumn{9}{|c|}{\textbf{GPT-4o}} \\
\hline
Zero-Shot 
& 66.37 & 40.04 & 56.73 & 22.39 & 46.91 & 19.36 & 58.67 & 30.62 \\

CoT+Few-Shot 
& \underline{86.09} & 51.54 & 71.04 & 31.53 & 55.14 & 28.31 & 73.88 & 40.93 \\

Multi-Agent 
& 85.72 & \underline{59.43} 
& \underline{73.17} & \underline{34.54} 
& \underline{58.38} & \underline{31.14} 
& \underline{75.06} & \underline{46.42} \\

\textbf{ChatModel} 
& \textbf{99.58} & \textbf{96.72} 
& \textbf{94.74} & \textbf{83.29} 
& \textbf{89.36} & \textbf{81.17} 
& \textbf{95.57} & \textbf{89.61} \\
\hline

\multicolumn{9}{|c|}{\textbf{Claude 3.7-Sonnet}} \\
\hline
Zero-Shot 
& 70.55 & 48.34 & 66.13 & 24.92 & 41.25 & 25.16 & 60.63 & 37.17 \\

CoT+Few-Shot 
& \underline{82.35} & \underline{60.23} 
& \underline{78.37} & 30.03 
& 47.87 & 27.39 
& \underline{70.88} & 44.91 \\

Multi-Agent 
& 75.41 & 57.75 
& 74.66 & \underline{34.31} 
& \underline{53.75} & \underline{32.78} 
& 68.50 & \underline{46.02} \\

\textbf{ChatModel} 
& \textbf{98.71} & \textbf{94.32} 
& \textbf{95.57} & \textbf{84.29} 
& \textbf{86.83} & \textbf{79.50} 
& \textbf{94.46} & \textbf{88.00} \\
\hline

\multicolumn{9}{|c|}{\textbf{DeepSeek-R1}} \\
\hline
Zero-Shot 
& 72.47 & 52.54 & 69.92 & 27.85 & 35.42 & 20.95 & 60.44 & 38.53 \\

CoT+Few-Shot 
& \underline{80.26} & \underline{57.23} 
& 72.43 & 31.50 
& 39.76 & 26.28 
& 66.27 & 43.24 \\

Multi-Agent 
& 77.62 & 54.29 
& \underline{73.51} & \underline{33.11} 
& \underline{44.52} & \underline{31.71} 
& \underline{66.56} & \underline{43.68} \\

\textbf{ChatModel} 
& \textbf{93.73} & \textbf{85.53} 
& \textbf{80.26} & \textbf{78.32} 
& \textbf{76.99} & \textbf{71.39} 
& \textbf{86.24} & \textbf{79.90} \\
\hline

\end{tabular}
\label{tab:performance_by_llm}
\end{table*}

\begin{figure*}[ht]
\centering
\includegraphics[width=1\linewidth]{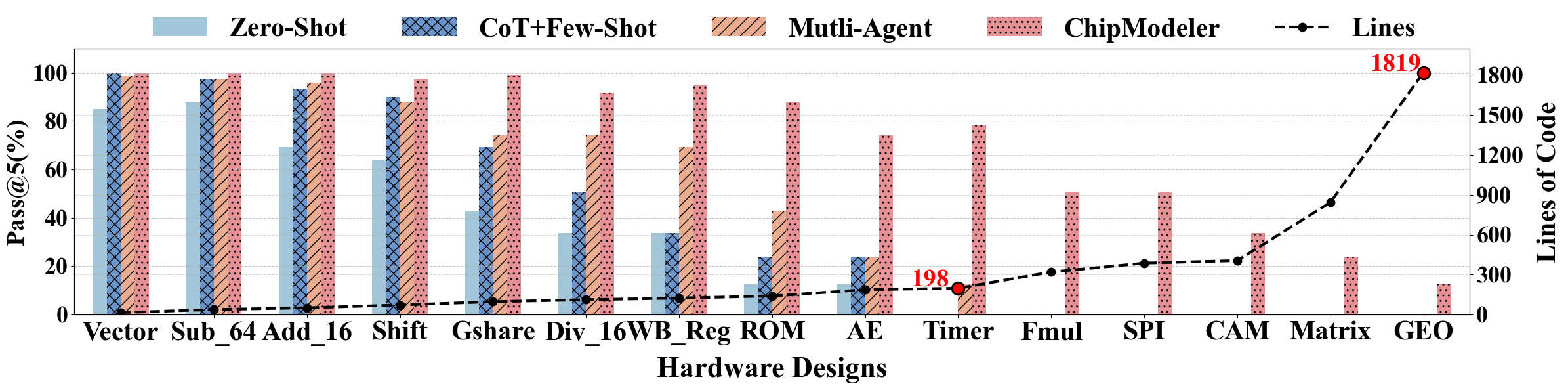}
\caption{Pass@5 Comparison of ChatModel and LLM-based Methods on SystemC Reference Model Generation for Specific Designs. The red dots indicate the maximum code generation capability of LLM-based methods and ChatModel, respectively.}
\label{fig:part_design}
\end{figure*}

\begin{figure*}[ht]
    \centering
    \begin{minipage}[t]{0.3\textwidth}
        \centering 
        \includegraphics[width=\linewidth]{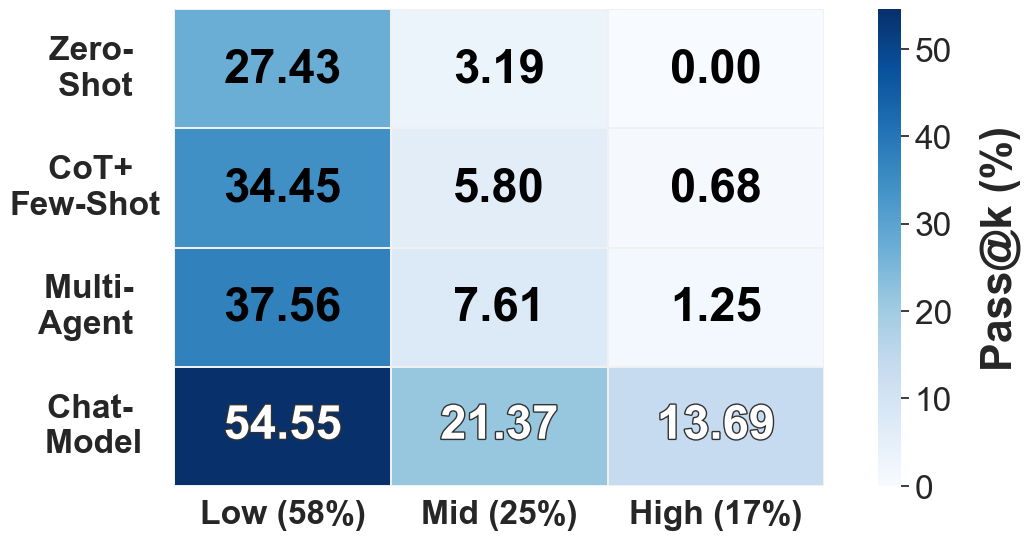}
        \caption{Pass@5 Results on Low, Medium, and High Scale Datasets with Different Methods.}
        \label{fig:category_results}
    \end{minipage}
    \hspace{0.01\textwidth}
    \begin{minipage}[t]{0.3\textwidth}
        \centering
        \includegraphics[width=\linewidth]{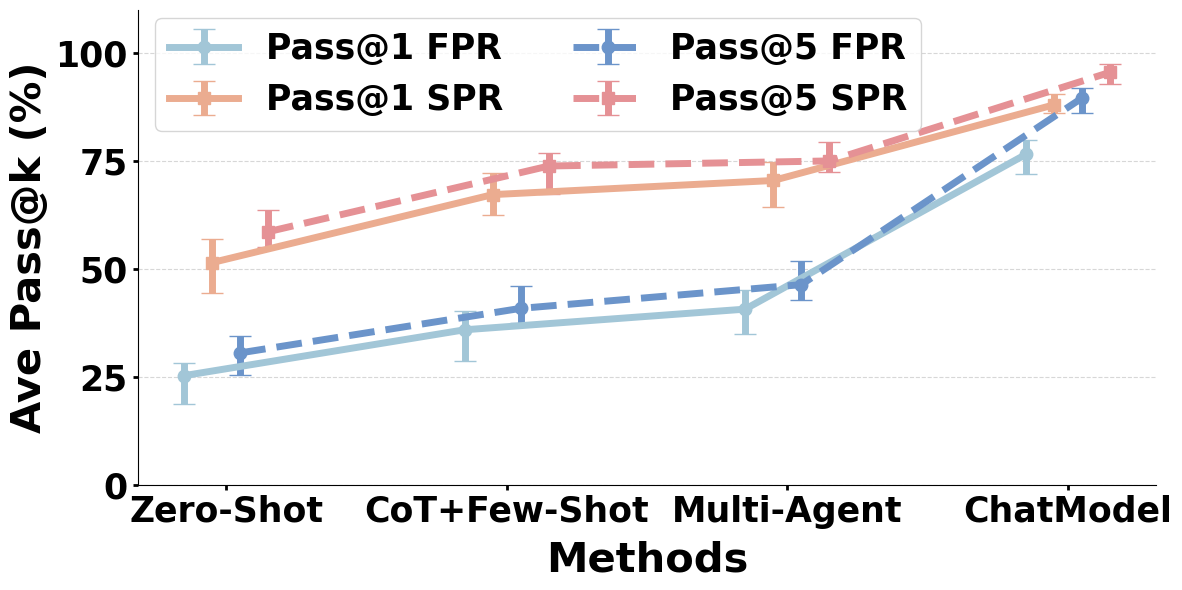}
        \caption{Performance and Stability Analysis of SPR \& FPR Pass@k (k=1,5) over Five Runs.}
        \label{fig:set_pass}
    \end{minipage}
    \hspace{0.01\textwidth}
    \begin{minipage}[t]{0.3\textwidth}
        \centering
        \includegraphics[width=\linewidth]{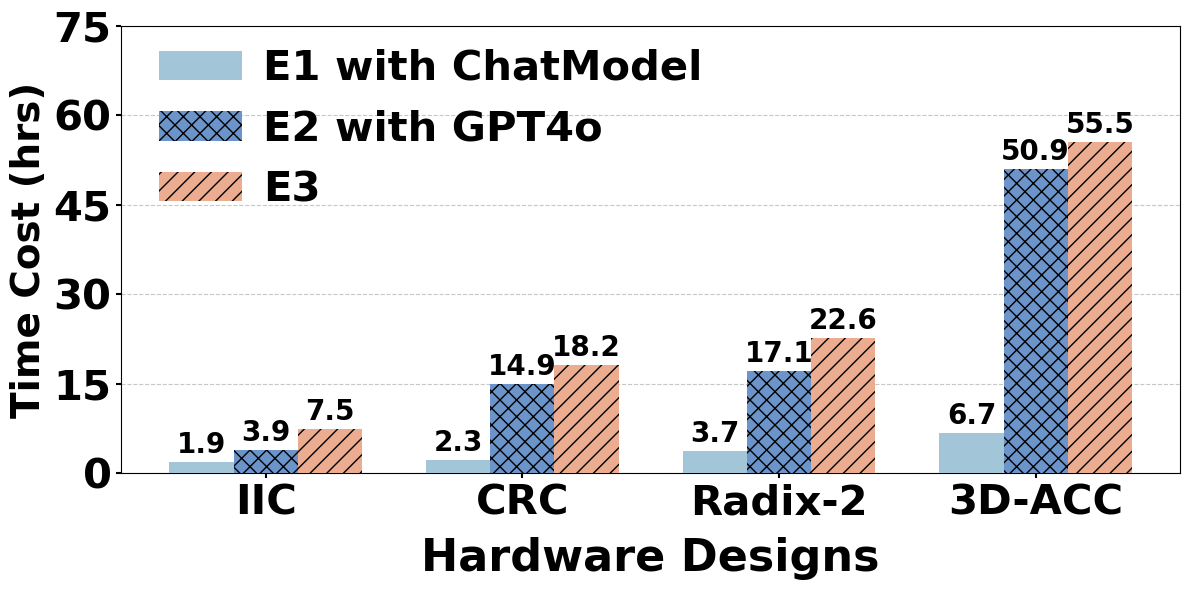}
        \caption{Time Cost Distribution with Different Methods.}
        \label{fig:Time_Evaluation}
    \end{minipage}
\end{figure*}

\begin{figure*}[ht]
\centering
\includegraphics[width=1\linewidth]{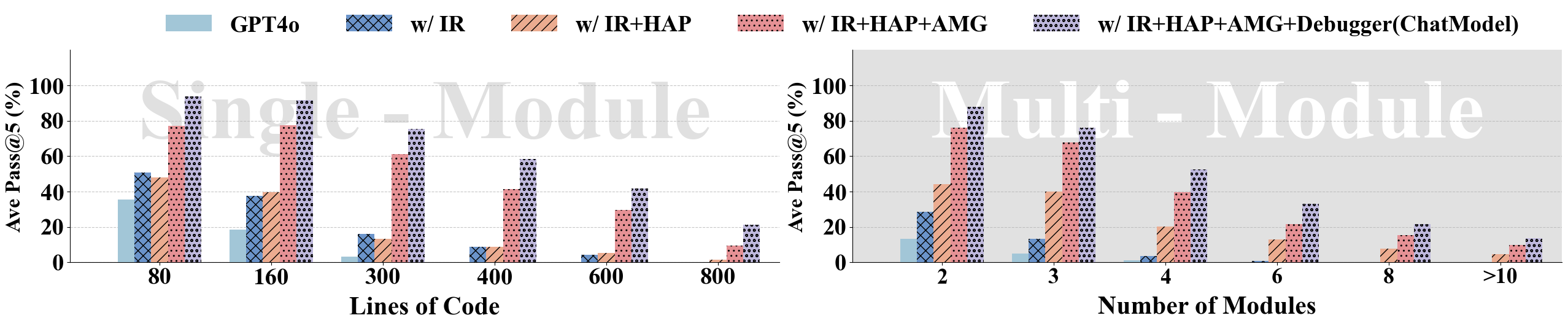}
\caption{Ablation Experiments: Evaluating the Impacts of Design IR, Hierarchical Adaptive Planning (HAP), Agile Model Generator (AMG), and Debugger in ChatModel.}
\label{fig:ablation}
\end{figure*}

As shown in Table~\ref{tab:performance_by_llm}, ChatModel consistently achieved the highest scores on both syntax and functionality tests across all evaluated LLMs and benchmarks, with an average Syntax Pass Rate (SPR) of \textbf{95.57\%} and a Functionality Pass Rate (FPR) of \textbf{89.61\%}.

Compared to the baseline FPR of the LLM-based methods, ChatModel demonstrates a performance improvement ranging from 36.22\% to \textbf{58.99\%}, significantly improving the ability of LLMs to generate accurate reference models. In Figure~\ref{fig:part_design}, we selected several representative examples from the three benchmarks, arranged in order of increasing complexity. As the complexity of the designs increases, the scores of the baseline methods decrease rapidly, whereas ChatModel consistently manages to complete these designs effectively. The figure demonstrates that ChatModel, compared to other LLM-based methods, generates complex designs of over 1800 lines and enhances design capability by \textbf{9.18$\times$}.

To better evaluate ChatModel's performance improvements at different generation scales, we group the designs by generation scale and compare their average Pass@5 rates in Figure~\ref{fig:category_results}.
Zero-Shot and other methods perform reasonably well on small-scale design tasks, but their effectiveness declines markedly with medium-scale and large-scale designs. In contrast, ChatModel consistently exhibits superior generative pass rates across datasets of all sizes. In comparison to the second highest Multi-Agent approach, ChatModel achieves enhancements of 1.45$\times$, 2.81$\times$, and \textbf{10.96$\times$} on the three test sets, respectively, highlighting its substantial advantage in handling complex design tasks.

Additionally, we performed five repeated tests on all benchmarks with various k values, as shown in Figure~\ref{fig:set_pass}. The results indicate that ChatModel not only enhances the ability of LLMs to generate reference models, but also significantly improves stability and robustness. Under the pass@1 condition, ChatModel achieves a 51.22\% improvement in FPR compared to the Zero-Shot baseline, with variability controlled at 7.75\%. This suggests that the majority of tests consistently pass the first time. Moreover, as the value of k increases, the improvement in the FPR increases to 58.99\%, while the variability decreases to 4.27\%.

\subsection{Agile Design and Verification Analysis}

To evaluate the acceleration provided by ChatModel in reference model design, we engaged three verification engineers with 2 to 3 years of experience who are familiar with SystemC. They independently developed and verified reference models for four designs of varying complexity and recorded their time costs. Engineer E1 utilized ChatModel for development, E2 employed GPT-4o for assistance, while E3 developed without the aid of LLMs. As illustrated in Figure~\ref{fig:Time_Evaluation}, ChatModel significantly improved development efficiency compared to E3, achieving an average speedup of \textbf{7.11$\times$}, with a range from 3.94$\times$ to 8.28$\times$.

\subsection{Ablation Experiment}

To further clarify the contribution of each component to the improvement in the generation capacity of ChatModel, we performed ablation experiments on designs of varying complexity. The entire experiment used GPT-4o as the original reference for comparison, and sequentially added four core elements of ChatModel: design IR, Hierarchical Adaptive Planning (HAP), Agile Model Generator (AMG), and debugger to observe the impact on generation performance.

Figure~\ref{fig:ablation} demonstrates that the incorporation of design IR significantly improves generative performance, with HAP further enhancing the capabilities of LLMs in handling large and complex designs. This improvement is attributed to the effective reduction of hallucinations in LLMs through the standardization of the specification, while HAP further mitigates the constraints imposed by token limitations. For more intricate designs, AMG effectively addresses the challenges of generating extended submodules and aligning interfaces. Furthermore, the introduction of a Debugger agent further strengthens the debugging capabilities of ChatModel.

\section{Conclusion}
\label {sec:con}
In this work, we introduce \textbf{ChatModel}, an agile framework for the generation and verification of reference models for general hardware. This framework effectively addresses the challenges faced in reference model design using LLMs through the standardization of design specifications and a structured Hierarchical Agile Modeling (HAM) process. We have also developed a debugger agent to effectively detect and correct errors in the generated models. 
Compared to traditional LLM-assisted reference model generation, our approach achieved a 58.99\% improvement in functional pass rate and expanded the generative scale of the reference model by 9.18X. In comparison to manually developed reference models, ChatModel demonstrates an enhancement in reference model generation and performance ranging from 3.94X to 8.28X. Our evaluation demonstrates the robust capability of ChatModel in generating complex reference models and further explores the transformative potential of LLMs in assisting reference model generation.

\newpage
{\small
\setlength{\bibsep}{0pt}
\bibliographystyle{plain}
\bibliography{references}
}

\end{document}